\documentclass[preprint2]{emulateapj}
\usepackage{natbib}
\bibpunct[; ]{(}{)}{,}{a}{}{;}

\shorttitle{Observations of Enhanced EUV Continua During an X-Class Solar Flare Using SDO/EVE}

\shortauthors{Milligan et al.}

\begin{document}

\title{Observations of Enhanced EUV Continua During an X-Class Solar Flare Using SDO/EVE}

\notetoeditor{the contact email is r.milligan@qub.ac.uk and is the only one which should appear on the journal version}

\author{Ryan O. Milligan\altaffilmark{1}, Phillip C. Chamberlin\altaffilmark{2}, Hugh S. Hudson\altaffilmark{3}, Thomas N. Woods\altaffilmark{4}, Mihalis Mathioudakis\altaffilmark{1}, Lyndsay Fletcher\altaffilmark{5}, Adam F. Kowalski\altaffilmark{6}, Francis P. Keenan\altaffilmark{1}}

\altaffiltext{1}{Astrophysics Research Centre, School of Mathematics \& Physics, Queen's University Belfast, University Road, Belfast, Northern Ireland, BT7 1NN}
\altaffiltext{2}{Solar Physics Laboratory (Code 671), Heliophysics Science Division, NASA Goddard Space Flight Center, Greenbelt, MD 20771, USA}
\altaffiltext{3}{Space Sciences Laboratory, UC Berkeley, 7 Gauss Way, Berkeley CA USA 94720-7450}
\altaffiltext{4}{Laboratory for Atmospheric and Space Physics, University of Colorado, Boulder, CO 80303, USA}
\altaffiltext{5}{School of Physics and Astronomy, SUPA, University of Glasgow, Glasgow G12 8QQ, UK}
\altaffiltext{6}{Astronomy Department, University of Washington, Box 351580, Seattle, WA 98195, USA}

\begin{abstract}
\noindent
Observations of extreme-ultraviolet (EUV) emission from an X-class solar flare that occurred on 2011 February 15 at 01:44~UT are presented, obtained using the EUV Variability Experiment (EVE) onboard the Solar Dynamics Observatory. The complete EVE spectral range covers the free-bound continua of \ion{H}{1} (Lyman continuum), \ion{He}{1}, and \ion{He}{2}, with recombination edges at 91.2, 50.4, and 22.8~nm, respectively. By fitting the wavelength ranges blue-ward of each recombination edge with an exponential function, lightcurves of each of the integrated continua were generated over the course of the flare, as well as emission from the free-free continuum (6.5--37~nm). The \ion{He}{2} 30.4~nm and Lyman-$\alpha$ 121.6~nm lines, and soft X-ray (0.1--0.8~nm) emission from GOES are also included for comparison. Each free-bound continuum was found to have a rapid rise phase at the flare onset similar to that seen in the 25--50~keV lightcurves from RHESSI, suggesting that they were formed by recombination with free electrons in the chromosphere. However, the free-free emission exhibited a slower rise phase seen also in the soft X-ray emission from GOES, implying a predominantly coronal origin. By integrating over the entire flare the total energy emitted via each process was determined. We find that the flare energy in the EVE spectral range amounts to at most a few per cent of the total flare energy, but EVE gives us a first comprehensive look at these diagnostically important continuum components.
\end{abstract}


\section{INTRODUCTION}
\label{intro}

The extreme-ultraviolet (EUV) portion of the solar spectrum is known to exert a significant influence on the Earth's upper atmosphere, particularly during periods of increased solar activity \citep{sutt06,qian10}. For example, fluctuations in the level of EUV flux received at 1~AU can generate changes in the ionospheric density. \cite{kane71} and \citet[and later \citealt{emsl78}]{donn78} quantified these changes using broadband SFD (sudden frequency deviation) ionospheric observations to show that the EUV flux from 1--103~nm (10--1030~\AA) rises and falls during solar flares in correlation with hard X-ray (HXR) emission, typically around 10--20~keV. The authors concluded that this EUV emission (which comprises free-free, free-bound, and bound-bound emission) originated in the solar chromosphere and was due to interactions with free electrons liberated by nonthermal particles accelerated during the flare's impulsive phase.

In addition to being the source of the most geoeffective emission during solar flares, the chromosphere is also where the bulk of a flare's energy is radiated \citep{neid89,huds92,huds06} as well as the source of the high-temperature plasma visible in coronal loops via the process of chromospheric evaporation (e.g. \citealt{mill06a,mill06b,mill09}). EUV (and UV) observations of the chromosphere are therefore an important diagnostic of the heating mechanism(s) responsible for driving increased emission at these wavelengths. Simulations by \cite{allr05}, using the RADYN code of \cite{carl94,carl95,carl97}, which modelled the chromospheric response to both electron beam heating and backwarming from XEUV (X-ray and EUV) photons, suggested that chromospheric emission is energetically dominated by various recombination (free-bound) continua, in particular, the Lyman, Balmer, and Paschen continua of hydrogen, and the \ion{He}{1} and \ion{He}{2} continua, as opposed to line (bound-bound) emission.

\begin{table*}[!t]
\begin{center}
\small
\caption{\textsc{\small{Wavelength Range, Time of Peak Emission, and Total Energy Radiated for each of the Processes Presented}}}
\label{eve_time_energy}
\begin{tabular}{lccc} 
\tableline
\tableline
\multicolumn{1}{c}{} &Wavelength Range &Time of Peak Emission				&Total Energy Radiated\\ 
\multicolumn{1}{c}{} &(nm)			 &Relative to GOES Peak (01:56 UT) 	& From 01:45--04:00~UT (ergs)\\ 
\tableline
Ly-$\alpha$ line		&116.6--126.6	 	&-0m38s         	&1$\times$10$^{30}$ \\
Free-free continuum		&6.5--35			&+4m52s		&8$\times$10$^{29}$ \\
GOES X-rays			&0.1--0.8	 		& 0m00s         	&5$\times$10$^{29}$ \\
Lyman continuum		&60--91.2          		&-2m38s		&4$\times$10$^{29}$ \\
He II 30.4~nm line    		&30.29--30.49  		&-0m18s		&3$\times$10$^{29}$ \\
He I continuum			&44--50.4			&+0m42s		&4$\times$10$^{28}$ \\
He II continuum		&21--22.8			&+5m42s		&1$\times$10$^{28}$ \\
\tableline
\normalsize
\end{tabular}
\end{center}
\end{table*}

Definitive observations of free-bound emission during solar flares have been scarce in recent years as many modern space-based instruments do not have the sensitivity, wavelength coverage, or duty cycle required to capture unambiguous continuum enhancements during flares. During the 1970's, chromospheric continuum observations were much more prevalent due to instruments onboard Skylab and the Orbiting Solar Observatory (OSO) series of satellites (e.g. \citealt{ziri75,lins76,mach78}). More recently, \cite{lema04} made a serendipitous detection of increased Lyman continuum from an X-class flare via scattered light in the Solar Ultraviolet Measurements of Emitted Radiation (SUMER) instrument onboard the Solar and Heliospheric Observatory (SOHO) satellite. \cite{chri03} also made a fortuitous detection of increased Lyman continuum emission during a stellar flare on an active M-type star with instruments onboard the Extreme Ultra-violet Explorer (EUVE). In addition, they presented tentative evidence for increased helium continuum emission blue-ward of the 50.4~nm recombination edge.  Understanding how different continua contribute to the overall energy of flares, including the depth of the atmosphere at which it is emitted and the mechanisms by which it is generated, are therefore crucial for corroborating solar flare models.

In this Letter we present new observations from the EUV Variability Experiment (EVE; \citealt{wood10}) instrument onboard the Solar Dynamics Observatory (SDO; the first mission from NASA's Living With A Star program) which show unambiguous, spectrally and temporally-resolved detections of the enhanced continua during the first X-class flare of Solar Cycle 24 which occurred on 2011 February 15. Here we focus on the free-free continuum, and on the free-bound continua of hydrogen and helium. EVE was primarily designed to measure the variations in the solar EUV irradiance on timescales from seconds to decades, and with high precision. However, our findings, in conjunction with X-ray observations from the Ramaty High-Energy Solar Spectroscopic Imager (RHESSI; \citealt{lin02}), highlight EVE's ability to help answer long-standing, fundamental questions on solar flare energetics, in particular, the structure and dynamics of the lower solar atmosphere and the origins of white light emission. Section~\ref{data_anal} describes the EVE data and their analyses, while Section~\ref{results} compares the EVE data with those from GOES and RHESSI to discuss the timing and energetics of the continuum emission and possible heating mechanisms. Section~\ref{conc} discusses the implications of these findings and their importance for future studies.

\begin{figure}[!t]
\begin{center}
\includegraphics[height=8.5cm,angle=90]{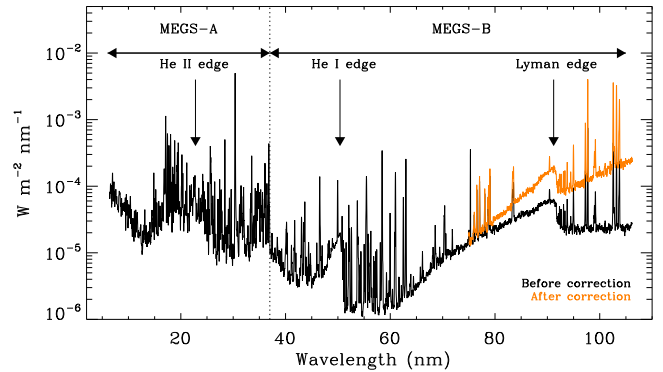}
\caption{A sample of the EUV spectra from 6.5--105~nm taken with both the MEGS-A and MEGS-B components of EVE prior to the X-class flare on 2011 February 15. The locations of the recombination edges at 22.8~nm, 50.4~nm, and 91.2~nm are noted. The data above 75~nm have been calibrated independently using TIMED/SEE data (orange curve; see text).}
\label{eve_spec}
\end{center}
\end{figure}

\section{EVE Observations and Data Analysis}
\label{data_anal}

The EVE instrument acquires full disk (Sun-as-a-star) EUV spectra every 10 seconds over the 6.5--37~nm (65--370~\AA) wavelength range using its MEGS-A (Multiple EUV Grating Spectrographs) component with a near 100\% duty cycle. Its MEGS-B component, which covers the 37-105~nm wavelength range, and MEGS-P component, a broadband (10~nm)\footnote[1]{Although the MEGS-P diode is 10~nm wide, more than 99\% of the detected emission is due to solar Ly-$\alpha$ after accounting for non-solar sources (e.g. geocoronal emission; \citealt{wood10}).} diode covering the Lyman-$\alpha$ line at 121.6~nm, only acquire data for three hours per day, and 5 minutes every hour to minimize instrument degradation. However, for one or two days per month, particularly if the Sun is active, MEGS-B and MEGS-P take data continuously to measure changes in the EUV irradiance  during the largest eruptive events. During one such campaign (2011 February 14--16), the first X-class flare of Solar Cycle 24 occurred (an X2.2 flare which began at 01:44~UT; SOL2011-02-15T01:56). Figure~\ref{eve_spec} shows sample EVE spectra from both MEGS-A and MEGS-B taken just before the flare occurred. Noted on the figure are the locations of the recombination edges of \ion{He}{2} at 22.8~nm, \ion{He}{1} at 50.4~nm, and the Lyman edge at 91.2~nm.

The current EVE data products (level 2, version 3) are validated up to 75~nm~with an accuracy of about 10\%. While the validation effort of the EVE irradiance is on-going, the uncertainty of the MEGS-B irradiance values in the 75--105~nm range is much larger, as much as an order of magnitude. However, for the purposes of this study, a cross-calibration with the Thermosphere Ionosphere Mesosphere Energetics and Dynamics/Solar EUV Experiment (TIMED/SEE; \citealt{wood05}) instrument was performed, which revealed a correction factor of 
\begin{equation}
9.15 - 0.167 \lambda +  0.000771 \lambda^{2},
\end{equation}

\noindent
where $\lambda$ is in nm. This correction factor is only valid for MEGS-B data in the range 75~nm$<\lambda<$91.2~nm taken during February 2011 and allows the absolute irradiance from the Lyman continuum to be measured. The corrected data are shown in orange in Figure~\ref{eve_spec}, and are not applicable to the \ion{C}{1} continuum, the tail of which is visible above 91.2~nm.

\begin{figure*}[!t]
\begin{center}
\includegraphics[height=14.5cm,angle=90]{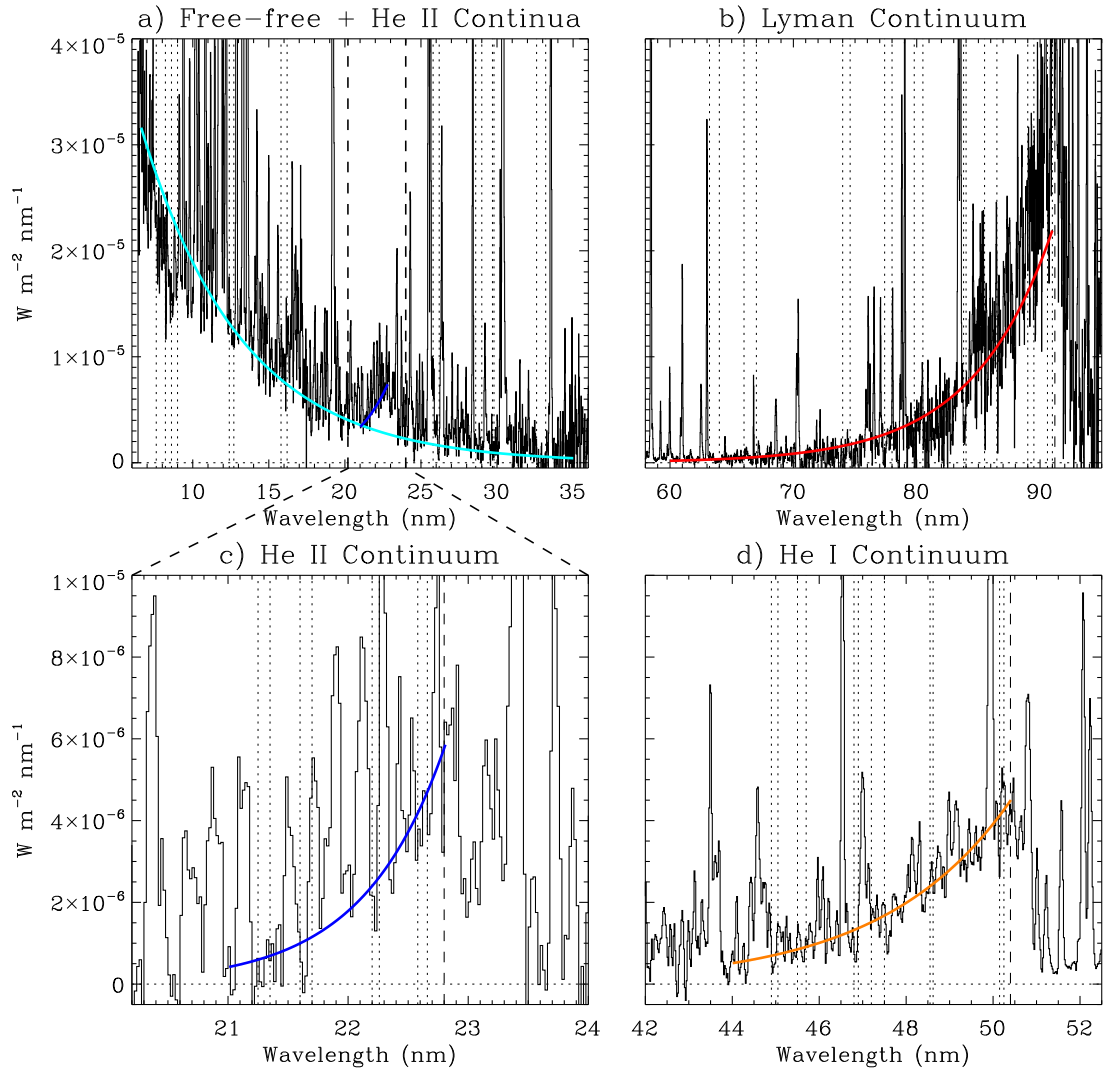}
\caption{Plots of the EUV emission observed by EVE near the SXR peak of the 2011 February 15 flare (01:55:32~UT), after subtracting out a pre-flare profile. a) Fits to the free-free continuum from 6.5--35~nm (cyan curve). The vertical dashed lines denote the wavelength range covered in panel c) while the dark blue curve represents the fit to the \ion{He}{2} continuum before subtracting the underlying free-free emission. b) Fit to the Lyman-continuum from 60--91~nm (red curve). The vertical dashed line denotes the recombination edge at 91.2~nm. c) The fit to the \ion{He}{2} continuum emission (dark blue curve) after subtracting out the underlying free-free emission. The vertical dashed line denotes the recombination edge at 22.8~nm. d) Fit to the \ion{He}{1} continuum (orange curve). The vertical dashed line denotes the recombination edge at 50.4~nm. In each panel, the vertical dotted lines show the portions of the spectrum predominantly free of any emission lines (based on synthetic spectra from CHIANTI) which were used to fit the continua. {\it See online material for a movie of these fits throughout the event.}}
\label{plot_cont_fits}
\end{center}
\end{figure*}

Figure~\ref{plot_cont_fits} shows portions of the EVE spectrum taken near the SXR peak of the flare (after subtracting out a 90-second averaged - from 01:00:12--01:01:42~UT - pre-flare profile) across the entire MEGS-A wavelength range (panel $a$), and around each of the three recombination edges (denoted by the vertical dashed lines in panels $b$--$d$). These data show that the free-free emission, which spans the entire MEGS-A wavelength range, and the free-bound emission blue-ward of each recombination edge were clearly enhanced during the flare. To estimate the absolute increase in continuum emission due to the flare, the data were fit with exponential functions to approximate the contribution from each continuum. In order to reduce the influence of the myriad emission lines superimposed on each continuum, a synthetic spectrum for each wavelength range was generated using CHIANTI, assuming its standard flare DEM (from \citealt{dere79}), coronal abundances, the ionization fractions of \cite{mazz98}, and an electron density of 10$^{11}$~cm$^{-3}$. Although the standard CHIANTI flare DEM is only relevant to the decay phase of flares, this synthetic spectrum was used merely as a guide to identify spectral regions which were free of any emission lines (at the EVE resolution) as denoted by the vertical dotted lines in Figure~\ref{plot_cont_fits}, rather than to infer any properties of the measured continua themselves. These regions were then averaged over and fit with the exponential functions. Fits to the spectrum near the SXR peak of the flare are denoted by the solid colored lines overlaid on each panel in Figure~\ref{plot_cont_fits} ({\it See online material for a movie of these fits throughout the event}).

\begin{figure*}[!t]
\begin{center}
\includegraphics[width=14.5cm,angle=90]{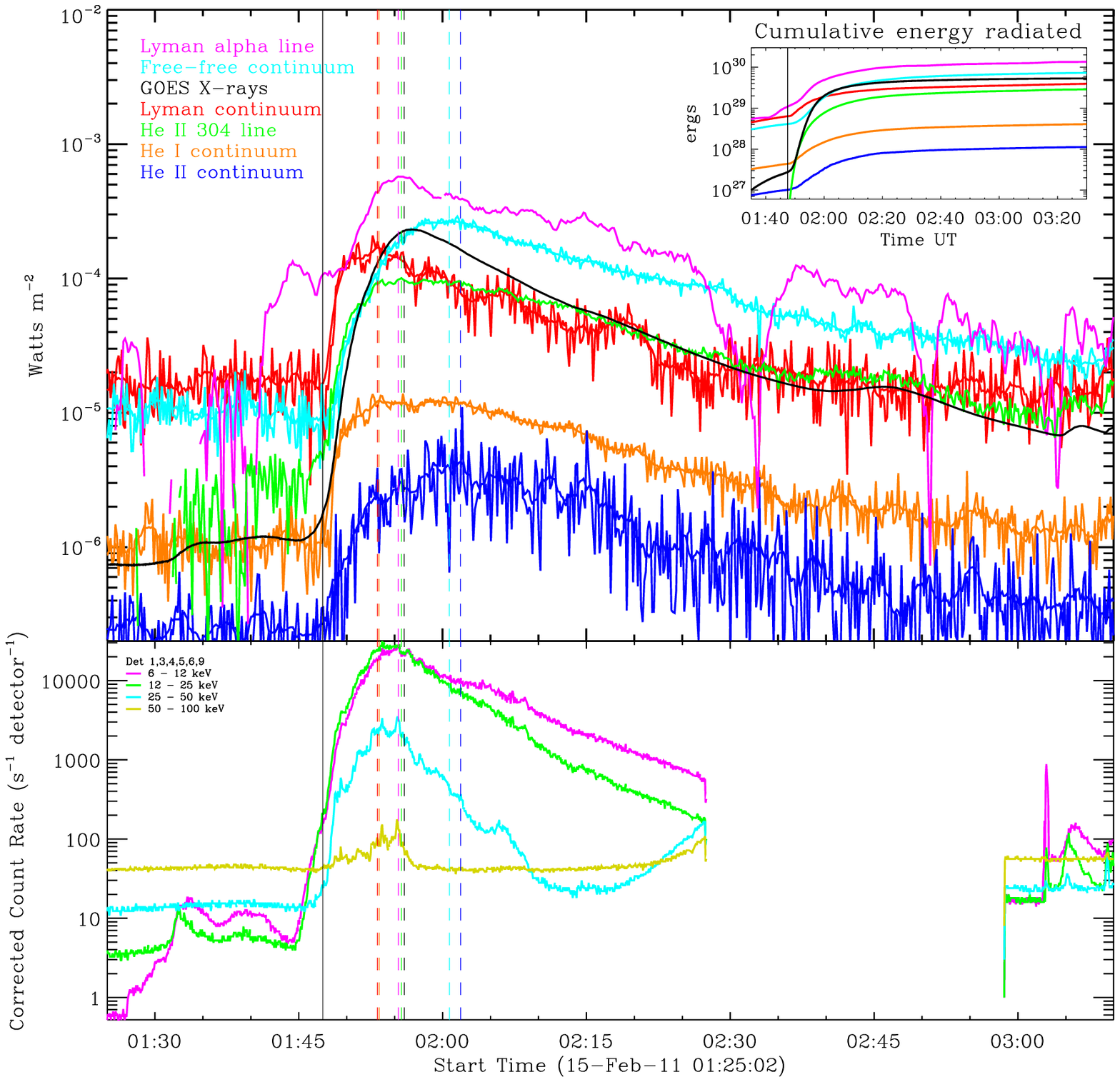}
\caption{Top panel: lightcurves of free-free, free-bound, and bound-bound emission as observed by EVE during the 2011 February 15 flare: Ly-$\alpha$ line - purple curve; free-free continuum - cyan curve; Lyman continuum - red curve; \ion{He}{2} 30.4~nm line - green curve; \ion{He}{1} continuum - orange curve; \ion{He}{2} continuum - blue curve. Note that the Ly-$\alpha$ channel (MEGS-P) contains large noise fluctuations, an example of which can be seen at about 01:44~UT. The vertical dashed lines in each panel denote the time of peak emission in each band. For each continuum, the peak times were derived from the smoothed profiles overlaid on each lightcurve. Also shown is the GOES 0.1--0.8~nm lightcurve (solid black line). The inset shows the cumulative energy radiated throughout the course of the flare by each process. Bottom panel: RHESSI 6--100~keV lightcurves of the same event. The vertical solid black line in each panel marks the approximate onset time of the continuum emission (01:47:30~UT).}
\label{eve_hsi_goes_ltc}
\end{center}
\end{figure*}

The free-free emission shown in panel $a$ is probably due to bremsstrahlung radiation from free electrons in the corona, unlike the free-bound emission from \ion{H}{0} and \ion{He}{0} which is chromospheric in origin. Superimposed on top of the free-free continuum, in addition to the multitude of emission lines formed between 10$^5$--10$^7$~K, lies the free-bound continuum of \ion{He}{2} (denoted by the blue curve shortward of the edge at 22.8~nm). Therefore, to accurately measure the contribution from the \ion{He}{2} continuum, the free-free contribution must be subtracted from the data before fitting the data between 21 and 22.8~nm. The fit to the free-free subtracted data is shown in panel $c$.

\section{Results}
\label{results}

The top panel of Figure~\ref{eve_hsi_goes_ltc} shows the total amount of irradiance emitted (in W~m$^{-2}$) by each continuum process, estimated by integrating under the exponential fits described in Section~\ref{data_anal} for each 10 second interval throughout the flare. Lightcurves of the \ion{He}{2} 30.4~nm (green curve) and Ly-$\alpha$ 121.6~nm (purple curve) lines are also included for comparison as these are believed to be significant radiators of the energy deposited by nonthermal electrons during flares \citep{allr05}. The GOES 0.1--0.8~nm X-ray lightcurve is shown as the solid black curve. This plot shows that the Ly-$\alpha$ line clearly dominates the radiative output, generating over 10$^{30}$~ergs throughout the event as indicated by the cumulative energy plot in the inset of the top right-and corner of the panel\footnote[2]{For MEGS-A and MEGS-B, 1~W~m$^{-2}$~nm$^{-1}$ at the Earth equals 2.812$\times$10$^{29}$~ergs at the Sun, for 0.02~nm wavelength bins and 10~s integrations, over a hemisphere of radius 1 AU. For the MEGS-P diode, 1~W~m$^{-2}$ = 1.406$\times$10$^{31}$~ergs also at 10~s integrations.} (see also \citealt{rubi09} for a summary of previous Ly-$\alpha$ flare observations). The wavelength and time ranges over which the irradiance was integrated are listed in Table~\ref{eve_time_energy}, along with the time of peak emission and the total amount of energy emitted via each process during the course of the flare.

The peak in the Ly-$\alpha$ lightcurve near 01:44 UT could be pre-flare enhancements, but is more likely due to measurement noise.  The EVE Ly-$\alpha$ channel measurement precision is only about 10\%, versus about 1\% for the other EVE channels.  Consequently, the EVE Ly-$\alpha$ measurements have more noise in their time series. In addition, this Ly-$\alpha$ channel is a photodiode versus a CCD for the MEGS channels, so the high energy particles in the SDO orbit cause spikes in the Ly-$\alpha$ data time series, versus just affecting a pixel or two on the CCD sensors. The data processing algorithms do attempt to make a correction for these spikes in the data time series, but not all of the spikes are successfully removed. The small peak of the Ly-$\alpha$ lightcurve near 01:44~UT could well just be effect of smoothing over one of these high energy particle spikes

Of the three free-bound continua studied, the Lyman continuum (red curve, Figure~\ref{eve_hsi_goes_ltc}) was found to be the most energetic. The time profile rose sharply at $\sim$01:47:30~UT (denoted by the solid black vertical line), synchronous with the 25--50~keV emission as observed by RHESSI (cyan curve, bottom panel), suggesting that the continuum emission was a result of hydrogen ions recombining with free electrons that were ionized in the chromosphere by nonthermal particles. The lightcurve also peaks around the time of the HXR bursts observed at 50--100~keV (vertical dashed red line). The total energy radiated by the Lyman continuum over the course of the flare was calculated to be 4$\times$10$^{29}$~ergs.

The \ion{He}{1} continuum showed a similar behavior to that of the hydrogen continuum, with an abrupt rise in step with the low energy HXRs, peaking with higher energy emission (vertical dashed orange line). However the total energy emitted was an order of magnitude lower at 4$\times$10$^{28}$~ergs.

The free-free continuum (cyan curve) began to rise at the same time as the Lyman continuum, but at a slower rate, and continued to increase after the Lyman continuum had peaked. The free-free emission reached its maximum at $\sim$02:01~UT; 5 minutes after the peak in GOES SXRs, and remained elevated above its pre-flare level for longer. As such, more energy was emitted by the free-free process compared to any of the free-bound continua over the same time period (8$\times$10$^{29}$~ergs). The \ion{He}{2} continuum, which is superimposed on top of the free-free continuum, also showed a similar gradual rise phase and a later time of peak emission, although was 40 times weaker. However, the measured timing may be due to inaccuracies in the fitting, as the \ion{He}{2} continuum is weak and noisy. It may also be biased by the underlying free-free emission which the helium emission had to rise above as the flare progressed.

Finally, the \ion{He}{2} 30.4~nm line also exhibited a shallower rise than that of the free-bound emission, but still peaked around the time of the HXR emission (vertical dashed green line). The total energy radiated by this line was comparable to that of the Lyman continuum at 3$\times$10$^{29}$~ergs.

\section{Conclusions}
\label{conc}

The findings presented in this Letter highlight the capability of the EVE instrument to determine increases in EUV continuum emission (both free-free and free-bound) during solar flares by giving the first comprehensive look at these diagnostically important emissions. From Figure~\ref{eve_hsi_goes_ltc} and Table~\ref{eve_time_energy} it is shown the free-free continuum emission dominated over the free-bound processes in terms of total energy emitted throughout the flare, although it peaked relatively later. By comparison the total energy emitted via emission lines (bound-bound processes) over the entire EVE spectral range (not including those listed in Table~\ref{eve_time_energy}) was 4$\times$10$^{30}$~ergs. Comparing these energies with the compilation by \cite{emsl05} it can be concluded that the total energy in the EVE range amounts to a few per cent of the total flare energy.

A comparison of the lightcurves derived from EVE with those from RHESSI indicate that the \ion{H}{1} and \ion{He}{1} continua are generated in the chromosphere as the hydrogen and helium ions recombine with free electrons ionized during the flare's impulsive phase. By contrast, the increase in free-free emission correlates well with GOES SXR emission suggesting that it is coronal in origin. This free-free emission may also be a significant contributor to images taken with the Atmospheric Imaging Assembly (AIA; \citealt{leme11}) instrument, also onboard SDO. AIA comprises 7 passbands that lie within the EVE/MEGS-A spectral range. As the free-free emission increases at shorter wavelengths, its contribution is likely to be greater in the higher temperature channels, such as 94~\AA~and 131~\AA~(see also \citealt{feld99,warr01}). 

These findings mark a significant advancement over the previous broadband (1--103~nm) observations of \cite{kane71} and \cite{donn78}, as now each continuum process can be individually resolved and determined as a function of time. As such, the continuum emission itself can be an important tool for diagnosing chromospheric heating when used in conjunction with X-ray observations from RHESSI, and numerical simulations such as the RADYN code of \cite{carl94,carl95,carl97}, which has been adapted by \cite{allr05} to include heating from nonthermal electrons and XEUV backwarming. Future EVE observations are therefore likely to provide a wealth of information on the structure and dynamics of the solar chromosphere, and its relationship to the flaring process.

\acknowledgments
This research was a result of several stimulating discussions between participants at a meeting on Chromospheric Flares held at the International Space Science Institute (ISSI) in Bern, Switzerland. ROM is grateful to the Leverhulme Trust for financial support from grant F/00203/X. HSH was supported by NASA under Contract NAS5-98033 for RHESSI. LF acknowledges financial support from STFC Grant ST/I001808 and the EC-funded FP7 project HESPE (FP7-2010-SPACE-1-263086).

\end{document}